\documentclass[a4paper,11pt]{article}
\pdfoutput=1 

\usepackage{jheppub} 

\usepackage[T1]{fontenc} 
\usepackage{amsfonts,amscd,amsmath, amssymb,graphicx,color}
\DeclareMathOperator\erf{erf}

\title{\boldmath Probing the Constituent Structure of Black Holes}

\author[a,b]{Lukas Gruending,}
\author[a]{Stefan Hofmann,}
\author[a,b]{Sophia M\"uller}
\author[a,b]{and Tehseen Rug}

\affiliation[a]{Arnold Sommerfeld Center for Theoretical Physics, 
	LMU-M\"unchen, Theresienstrasse 37, 80333 M\"unchen, Germany}
\affiliation[b]{Max-Planck-Institut f\"ur Physik,
	F\"ohringer Ring 6, 80805 M\"unchen, Germany}

\emailAdd{Lukas.Gruending@physik.uni-muenchen.de}
\emailAdd{Stefan.Hofmann@physik.lmu.de}
\emailAdd{Sophia.X.Mueller@physik.uni-muenchen.de}
\emailAdd{Tehseen.Rug@physik.lmu.de}

\abstract{Based on recent ideas, we propose a framework for the description of 
         black holes in terms of constituent 
         graviton degrees of freedom. Within this formalism a large black hole can be understood
         as a bound state of $N$ longitudinal gravitons. In this context black holes are 
         similar to baryonic bound states in quantum chromodynamics (QCD)
         which are  described by fundamental quark 
         degrees of freedom. 	As a quantitative tool we employ a quantum bound state description originally developed 
	in QCD that allows to consider black holes in a relativistic Hartree--like framework.
	As an application of our framework we calculate
          the cross section for scattering processes
	between graviton emitters outside of a Schwarzschild
	black hole and absorbers in its interior, that is gravitons. 
	We show that these scatterings allow to directly extract structural
	observables such as the momentum distribution of black hole constituents.}

\begin{document} 
\maketitle
\flushbottom

\section{Introduction}
\label{sec:intro}

In general relativity, the complete gravitational collapse of a spherical symmetric body 
results in a Schwarzschild black hole.
Based on the asymptotic flatness of the Schwarzschild solution, the 
black hole is fully characterized by the total mass. This allows to interprete the 
Schwarzschild metric in terms of the exterior gravitational field of an
isolated body. 
Duff \cite{Duff} showed that the Schwarzschild solution can be
obtained by resumming infinitely many tree--level scattering processes
involving weakly coupled gravitons and the black hole as an external
source on Minkowski space--time. Therefore, the exterior of
 a Schwarzschild black hole admits both, a geometrical and
  a quantum mechanical description based on the $S$--matrix.

In our opinion, the luxury of friendly coexisting descriptions ends at the event horizon
of the black hole. The reason can be understood as follows: the standard semi-classical
treatment of Hawking radiation inevitably leads to non-unitary time evolution in the sense
that pure states evolve into mixtures. This is known as the
information paradox which is true for arbitrarily large black holes
(excluding the possibility of remnants). In particular, this suggests that a resolution to this
problem could be insensitive to the details of a UV completion of gravity.
In contrast any sensible effective quantum field theory on
flat space--time should preserve information by default.
Let us stress already at this point
that we do not claim that quantum field theory (QFT) on curved space--time is not valid.
Rather it describes an idealized semi--classical situation which might miss
quantum effects which could lead to purification of Hawking radiation.

In this article, we want to explore the possibility that QFT on flat space--time
is fundamental, even for the description of black hole interiors.

The situation is somewhat analogous to the status
quo of the proton around the advent of quantum chromodynamics. 
The mass, spin and electrical charge of the proton were known.
Mass and spin are related to the Casimir invariants of the Poincare group,
i.e.~to the isometries of Minkowski space--time. 
Low--energy effective 
model building allowed to study hadron reactions at energies sufficiently low
to neglect the internal structure of the participating hadrons. 
Protons and hadrons in general, however, do not enter in the Hamiltonian 
of quantum chromodynamics, albeit they are part of its spectrum, 
just not as elementary degrees of freedom, but as non--perturbative bound states.
Understanding the internal structure of hadrons in terms of its (asymptotically) 
perturbative constituents becomes a formidable problem. 

The charge radius of protons serves as a working analogue to the 
Schwarzschild radius. The charge radius sets the average length scale for confining
color within protons. Hadrons in general cannot leak color, and 
chromodynamics can only be studied if probes are employed that can resolve length scales
smaller than the charge radius. This can be achieved experimentally in deep inelastic
scattering processes involving leptons emitting virtual photons and protons absorbing them. 
The only information observed is the recoil of the emitter. This information suffices to 
reconstruct the proton interior in terms of so--called structure functions. 
Outside the confining proton, 
questions can be answered using perturbation theory. Once the virtual photon has 
been absorbed by the proton, it probes the interior of a strong bound state. 
The interior structure of the proton depends on non--perturbative features of quantum chromodynamics,
and cannot be fully described by means of perturbation theory. While asymptotic freedom allows
a perturbative description for individual interactions at sufficiently small distances, 
confinement is a non--perturbative effect. It is also important to mention that confinement
is a priori not due to collective effects, where all constituents create an effective 
potential for each single constituent that would be responsible for confinement.
Nevertheless, Shifman et al. \cite{Shifman,Shifman1} showed that questions pertaining to the internal structure
of hadronic bound states can be formulated in a mean--field language, and the formulation
is as close as it gets to a relativistic Hartree--like approximation.
Shifman et al. postulated the existence of a non--perturbative ground state that supports 
the creation of bound states when an auxiliary current operates on it. The main difference
to the perturbative vacuum is that it allows for quarks and gluons to condense.
In turn, condensates of quarks and gluons parametrize the a priori unknown ground state. 
This way, non--perturbative effects are mapped to the details of this ground state such that 
generic observables factorize in perturbative (calculable) and non--perturbative (parametrized) pieces.    

The main question we wish to pursue in this article is whether the interior
 of Schwarz\-schild black holes admit a similar quantum bound state description. 
 If a quantum mechanical description of the 
black hole interior is at all feasible, it has to involve non--perturbative aspects
of the quantum theory. And these aspects need to quantify the difference for a free field
to evolve in the exterior or the interior of a Schwarzschild black hole. 
The very fact that the evolution is different motivates a quantum bound state description of 
the black hole interior.

Recently a description of black holes as quantum bound states of $N$ weakly
interacting constituents has been suggested by Dvali and Gomez \cite{Dvali, Dvali1}. 
At first sight, the complementary description suggested here seems difficult to
achieve, because black holes are bound states of $N\gg 1$ constituents. 
However, 't Hooft showed that large--$N$ systems can be blessed with simple 
scaling laws, giving rise to e.g. planar dominance \cite{tHooft}. 
In his original work, 't Hooft considered the case of a $SU(N)$ gauge theory
for $N\rightarrow\infty$. Witten considered heavy baryons in this theory,
which consist of $N$ constituent quarks in the fundamental representation 
of the gauge group, assuming that  $SU(\infty)$ is a confining gauge theory \cite{Witten}.

He showed that a diagrammatic approach to bound state properties (beyond the 
parton--level description) seems to indicate a bad large--$N$ limit. 
This lead to the observation that the large--$N$ behavior of this theory 
is only sensible provided the baryon mass scales as $M_\mathcal{B}\propto N$.
In extending these kinematical considerations towards a dynamical description,
Witten employed a Hartree approximation and restricted his work to heavy baryons. 

Notice that the assumption of confinement in $SU(\infty)$ was needed to 
ensure a proper bound state spectrum 
in terms of color singlet hadrons. In general, the microscopic origin
of the bound state spectrum, however, is not related to confinement or asymptotic
freedom in the UV. In particular, the use of mean--field techniques does
not depend on the precise nature of a large--$N$ system under consideration.
Consider for example an atom with many electrons.
In this case, constituents interact according to quantum electrodynamics.
Therefore, no confining mechanism in such systems is at work.
Nevertheless, as soon as the number of bound state constituents is large enough,
mean--field techniques based on the Hartree ansatz can be succesfully employed.

Witten \cite{Witten} demonstrated that the large--$N$ nature of the bound states lead to enormous simplifications 
as compared to real QCD. The reason for this is the emergence of a new expansion parameter,
$1/N$, i.e. the inverse number of bound state constituents. As explained before, the power 
of the large--$N$ logic is not restricted to asymptotically free gauge theories. In fact, for any
generic system composed of a large number $N\gg 1$ of individually weakly coupled quanta 
an expansion in $1/N$ can be employed.
Another important example is given by Bose-Einstein condensates consisting of $N$ bosons. 
In that case the Bogoliubov approximation amounts to a truncation of a $1/N$ expansion at 
leading order.
Thus, the simplifications rooting in the nature of large--$N$ systems seem to be generic, no matter
what the underlying reasoning of bound state formation or condensation is. 
The reason for this property can eventually be traced back to fact that all these systems allow 
for a mean field description.

In practice, such a description is feasible technically in non--relativistic systems.
However, as far as relativistic systems are concerned, there has not been much progress.
An implementation of the Hartree idea in such a situation, however, is of fundamental importance.
Applications include for example, large--$N$ baryons consisting of light quarks
or black holes in the picture of \cite{Dvali}.
In principle a mean--field approach is given in terms of a consistent truncation
of the Schwinger--Dyson equations. Technically, finding solutions to the truncated system,
is in general a very demanding task. Therefore, it is desirable, to find a different implementation 
of the mean--field idea in relativistic field theory. At a practical level,
this could, for example, amount to a parametrization of complicated microscopic phenomena 
in terms of phenomenological parameters. As such, these parameters can be effectively
interpreted as coarse--grained observables.

Actually, first steps in that direction were proposed
in \cite{Hofmann}. There a general framework addressing
relativistic mean--field questions involving bound states was
presented and developed.
For example, in the context of gravitational bound states, the only input of the ideas of \cite{Dvali}
is that black holes might be understood as large--$N$ bound states of longitudinal gravitons.
Scaling relations or other results based on the estimates given there are not assumed in the
formalism of \cite{Hofmann}
at any point. On the contrary, the field theoretical apparatus presented in this work
explicitly allows to derive such scalings.

The framework employs weakly
coupled constituents immersed in a complicated ground state
filled with constituent condensates. These condensates parametrize the strong collective effects 
discussed above and correspond to normal--ordered contributions
appearing when using Wick's theorem. Thus, the condensates create 
an effective coarse--grained potential
in which individual particles propagate. This construction is similar to the one used in sum rule calculations
in non--perturbative QCD \cite{Shifman, Shifman1} as discussed above.
The difference is that the strong effects in QCD are due to confinement
while in our case the ground state itself represents a mean field source, suggesting a relativistic
Hartree framework to address non-perturbative questions. 

As mentioned before the techniques developed in \cite{Hofmann}
are completely general and apply to both, gravitational and non--gravitational
relativistic bound states consisting of many quanta.

Since we are interested in bound state properties of Schwarzschild space--times,
we will however focus on spherically symmetric purely gravitating sources in this article. 
In what follows black holes are treated on the same footing as other gravitating bound states. 
The only difference is that the radius of the black hole is equal to the Schwarzschild radius.
Questions concerning thermality and entropy will be addressed in future work\footnote{The 
construction described in Section \ref{ACD} could, for example, be generalized to density
operators describing the statistical properties of bound states.}.

A priori, quantum bound states representing black holes are unknown.
Nevertheless, structural information about black holes understood as
graviton bound states can be extracted
from kinematical states that store quantum numbers
and isometries pertaining to black holes. 
This is a common procedure in quantum chromodynamics 
where, for example, the pion decay constant
is defined as the overlap between a two--quark state 
and the a priori unknown pion state ~\cite{Shifman, Shifman1}.

Our aim is to provide a quantitative formulation of these ideas.
Having developed the theoretical framework in \cite{Hofmann},
we will focus on applications within S-matrix theory in this article.

Black hole interiors, considered as quantum compositions of many constituents,
can be probed using virtual gravitons. 
Albeit the probes are absorbed by black hole constituents inside of the black hole,
 information about the constituent distribution
inside the bound state can be extracted  
from the probe emitter far away from the black hole. 

In this article, we use this framework to construct observables like
cross--sections. Furthermore, we study scattering processes that resolve
the constituent structure of quantum bound states associated with black holes.
In order to resolve the constituent composition of the black hole interior,
a virtual graviton has to be emitted nearby the black hole horizon
and be absorbed by the black hole. In such a inelastic scattering process,
the horizon is the boundary separating the perturbative vacuum 
in the black hole exterior from the non--perturbative ground state in the interior. 
The information concerning the constituent distribution inside the black hole  
can be extracted from the scattering angle between the emitter asymptotics. 
This would allow for a complementary description of the black hole interior based on 
observables which can be measured by an outside observer. 

In Section \ref{ACD} the auxiliary current description is presented as the key concept
allowing to represent quantum bound states by kinematical states which
carry structural information. After reviewing the general construction based on \cite{Hofmann}
we will specialize our reasoning to spherically symmetric gravitating bound states.
In Section \ref{BHS} to \ref{CD} we present our computation of the cross--section of
scattering of scalars on the black hole. We show that the result can be written in terms 
of distribution functions of gravitons inside the black hole. 
In particular, Section \ref{BHS} is devoted to the description of black holes as absorbers of virtual gravitons.
This description is promoted to absorption processes compatible with the S-matrix framework
in Section \ref{CO}. Section \ref{CR} offers an interpretation of these absorption processes in
terms of constituent observables, which in turn offers the possibility to extract 
black hole constituent observables from scattering experiments.
Section \ref{AP} complements this analysis by uncovering the analytical structure 
underlying the absorption of virtual messengers by black holes.
In Section \ref{CD} we present the graviton distribution at the parton level.
We want to stress, that our aim in this article is to show that
cross sections can be described in terms of constituent quanta.
Practically, this implies that the cross section is parametrized
by a non--perturbative quantity, the graviton distribution function.
This function should be measured in a real experiment.
Predictivity of our techniques would then follow from the renormalization
group evolution of that distribution function. This
is exactly the same as the analogous situation in QCD.
The question of renormalization group flows, however,  is 
left for future work.


\section{Auxiliary current description}\label{ACD}
Within the perturbative framework suitable for describing scattering processes
there is no dynamical constituent representation of bound states based on 
elementary degrees of freedom. This, however, does not exclude a sensible
representation of a bound state.

At the kinematical level all quantum states are identified by their quantum numbers.
These quantum numbers should be in accordance with the intrinsic symmetries at work 
(such as gauge symmetries), 
and with the isometries characterising bound states in Minkowski space--time. 
Furthermore, the state has to be characterised according to isometries of Minkowski (Casimir operators
of Minkowski). Including all these quantum numbers, collectively denoted as $\mathcal{L}$, leads to
a complete kinematic characterisation of the bound state in question. Let us, for example, consider 
a proton. Kinematically a proton must be a colour neutral state with the correct quark content
to ensure the correct electric charge and isospin corresponding to gauge quantum numbers.
A further characterisation of the state is given by the mass and spin which are the eigenvalues 
of the corresponding Casimir operators of Minkowski.
In the following, this construction will be demonstrated for the Schwarzschild case.

We assume a unique non--perturbative ground state $|\Omega\rangle$ 
which supports all quantum numbers  $\mathcal{L}$ in the bound state spectrum.
In particular, bound states can be created using appropriate auxiliary currents $\mathcal{J}$
acting on $|\Omega\rangle$. These currents should contain the field content associated with
the bound state at hand. For example the current for the $\rho$--meson is given by 
$\mathcal{J}_{\rho}^{\mu}=1/2 (\bar{u} \gamma^{\mu}u-\bar{d} \gamma^{\mu}d$) \cite{Shifman, Shifman1}, where $u$ and $d$
are the up and down quark fields, respectively. Notice that this current has the correct isospin, charge and
colour quantum numbers to represent a $\rho$--meson. This ensures that the overlap with the true state of the
$\rho$--meson is non--vanishing, thus allowing to express the true state $|\rho\rangle$ in terms of
$\mathcal{J}_{\rho}^{\mu}$. In our case the current should be composed
out of $N$ gravitons. We will come back to the explicit form of the current later in this section.

First we derive the representation of a generic bound state $|\mathcal{B}\rangle$ in Minkowski space--time
with quantum numbers encoded in $\mathcal{J}$. For that purpose consider

\begin{eqnarray}
\langle\Omega|\mathcal{J}(x)|\mathcal{B}\rangle&&=\int \frac{\rm{d}^4P}{(2\pi)^4} \mathcal{B}(P,\mathcal{Q})\rm{e}^{-\rm{i}Px}\underbrace{\langle\Omega|\mathcal{J}(0)|P\rangle}_{\Gamma_{\mathcal{B}}}.
\end{eqnarray}

Here we inserted a complete set of momentum eigenstates $|P\rangle$ and used translational
invariance of the vacuum state. Note that the matrix element on the right hand side
defines a non--trivial decay constant $\Gamma_{\mathcal{B}}$. As mentioned before this construction
is analogous to definitions of decay constants in the framework of QCD \cite{Shifman, Shifman1}.  $\mathcal{B}(P,\mathcal{Q})$
is the wave function of the bound state in momentum space carrying information about possible gauge
 quantum numbers and isometries encoded in $\mathcal{Q}$. Thus we identify the complete set of quantum 
 numbers $\mathcal{L}=\{P,\mathcal{Q}\}$.

Expanding $|{\mathcal{B}}\rangle$ and $\mathcal{J}(x)|\Omega\rangle$ separately in momentum eigenstates
and using the definition of $\Gamma_{\mathcal{B}}$ we arrive at the auxiliary current representation of an
arbitrary quantum state\footnote{Later, for example, we will explain how this construction reduces to the Lehmann--Symanzik--
Zimmermann formula in the context of perturbative $S$--matrix theory.} and, in particular, a bound state

\begin{eqnarray}
	|\mathcal{B}\rangle 
	&=&
	\frac{1}{\Gamma_\mathcal{B} }\int {\rm d}^4 P \; \mathcal{B}(P,\mathcal{Q}) \int \tfrac{{\rm d}^4x}{(2\pi)^4}
	\; {\rm e}^{{\rm i}P\cdot x} \mathcal{J}(x) |\Omega\rangle
	\; .
	\label{bhstate}
\end{eqnarray}	

Notice that $\mathcal{B}(P,\mathcal{Q})$ localizes the information
encoded in $\mathcal{J}$ in $|\mathcal{B}\rangle$	. It is intrinsically non--perturbative
and can be related to the distribution of gravitons inside the bound state
as we will discuss in more detail later.

Since in our construction spherically symmetric, gravitating sources are understood
as bound states on flat space--time, also here one characterisation is given
by the Casimir operators of Minkowski $P^2=-M_{\mathcal{B}}^2$.  
Thus our generic derivation applies here as well\footnote{
Note that on-shell we have $P^2 = -M_{\mathcal{B}}^2$. 
Thus, four-dimensional momentum integration
is only chosen for convenience. In particular, 
$\mathcal{B}(P,\mathcal{Q}) = \delta^{(1)}(P^2+M_{\mathcal{B}}^2) \hat{\mathcal{B}}(P,\mathcal{Q})$
with $\hat{\mathcal{B}}(P,\mathcal{Q})$ the on-shell wave function of the state $|\mathcal{B}\rangle$.}.
Here $M_{\mathcal{B}}$ is the bound state's mass.
Since we consider Schwarzschild black holes there are no gauge quantum numbers
associated to it. Therefore, in this discussion we do not need to take gauge quantum numbers
into account\footnote{These could be implemented easily, however. In the case of
electrically charged black holes one could choose the current in such a way that it contains $U(1)$ fields 
as well as gravitons.}.

Let us know specify the current $\mathcal{J}$ for the question at hand.
The current $\mathcal{J}$ carries isometry information of black hole quantum bound states 
appropriate for the kinematical description. Note that these isometries 
are not due to any geometrical concept. Instead, they are a consequence 
of the explicit breaking of certain Lorentz symmetries in the presence of bound states.

Black holes can be modelled using bound states of $N$ gravitons by means of
the following local, composite operator,
\begin{eqnarray}
	\mathcal{J}(x)
	=
	\mathcal{M}(\underbrace{h,\dots,h}_{N})(x) \;.
\end{eqnarray}	
Here $\mathcal{M}$ denotes a Lorentz covariant tensor
coupling $N$ gravitons $h$ in accordance with the bound state isometries 
archived in $\mathcal{Q}$.
For simplicity, we displayed only graviton 
couplings, but other degrees of freedom can be included in the current description. 
In fact, this will be necessary when gravitons are coupled to other fields. 
The bound state isometries are represented via their associated   
symmetry generator $\mathcal{K}$. In order for the state $|\mathcal{B}\rangle$ 
to respect them, they are realised as 
local symmetries of the currents, i.e.~$[\mathcal{J},\mathcal{K}]=0$. 
This implies the invariance of the state $\mathcal{J}(x)|\Omega\rangle$ 
under the action of  $\mathcal{K}$.

The local symmetry condition leads to a differential equation determining 
the space--time dependence of the current.
In our case, the collection of isometries includes generators
corresponding to temporal homogeneity and spatial isotropy. 
We find $h=f(r)$ with $r$ denoting the Euclidean distance, so
$\mathcal{J}=\mathcal{M}(h,\dots,h)(r)$. The coupling tensor $\mathcal{M}$
is not further constrained. The simplest auxilliary current is given by  
$\mathcal{J}=({\rm tr} h)^N(r)$. For notational simplicity we represent 
the bound state gravitons by massless scalars, $\mathcal{J}=\Phi^N$.  
This is completely justified at the partonic level,
where gravitons are non-interacting.
The reader is referred to \cite{Hofmann} for more details concerning
the choice of the current at this level of accuracy.

Any Observable $\mathcal{O}$ associated with black hole structure 
is subject to the isometries stored in $\mathcal{Q}$.
Using (\ref{bhstate}), Ward's identity leads to 
\begin{eqnarray}
      0=\langle\mathcal{B}|\partial_\mu j^\mu | \mathcal{B}\rangle
      =\langle\mathcal{B}|\delta\mathcal{O} | \mathcal{B}\rangle
      =\delta\langle\mathcal{B}|\mathcal{O} | \mathcal{B}\rangle.
      \label{ward}
\end{eqnarray}
Here, $j$ denotes a conserved current 
corresponding to an isometry (this should not be confused with the auxiliary current $\mathcal{J}$).
In practice, (\ref{ward}) implies that observables can be calculated
in a fully Lorentz covariant way and 
it suffices to impose the symmetry constraints in the end. 

Although the auxiliary current description is transparent, it is worth appreciating
its simplicity when applied to free states 
$|\chi\rangle=|{\bf k},\mathcal{Q}\rangle$:
\begin{eqnarray}
	a^\dagger({\bf k},\mathcal{Q}) |0\rangle
	&=&
	\Gamma_\chi^{\; -1}\int\tfrac{{\rm d}^3x}{(2\pi)^3} \; {\rm e}^{{\rm i}k\cdot x}
	\mathcal{J}(x) |\Omega\rangle
	\; ,
\end{eqnarray}	
where $k$ is on--shell, ${\bf k}$ is the particles three momentum and $|0\rangle$ denotes the perturbative vacuum state. 
In the auxiliary current description, excitations of the perturbative vacuum 
are generated by acting with the auxiliary current on the perturbative vacuum on a spatial slice at an arbitrary time. 
The current simply reduces to the field operator creating a scattering state from the vacuum.
For example, $\mathcal{J}(x)=\chi(x)$ for a single particle scalar scattering state.
Since $k$ is on--shell, an ingoing scattering state is given by
\begin{eqnarray}
	\label{in}
	|{\bf k},\mathcal{Q}, {\rm in}\rangle
	&=&
	\tfrac{2\pi}{{\rm i}\Gamma_\chi k^0({\bf k})}
	\int \tfrac{{\rm d}^4x}{(2\pi)^4} \; {\rm e}^{{\rm i}k\cdot x}
	D(x) \mathcal{J}(x) |\Omega\rangle \; ,
\end{eqnarray}	 
with $D(x)$ denoting the equation of motion operator associated with $|\chi\rangle$.
Note that in (\ref{in}) a boundary term, as well as a 
term that would lead to a disconnected contribution in the scattering matrix have been dropped. 

Hence, the auxiliary current description reduces to the 
famous Lehmann--Symanzik--Zimmermann reduction formula when applied to scattering states. 
This suggests a unified framework for scattering processes involving constituent and asymptotic states.

\section{Black hole structure}\label{BHS}
In this section the non--geometrical concept of black hole structure will be introduced.  

Quantum field theory allows for two distinct source types, external and internal sources,
referring to the absence and presence of sources in the physical Hilbert space, respectively. 
They are, however, not on equal footing, since external sources approximate
a more fundamental description solely involving internal sources.
Clearly, an external source is structureless, while 
small--scale structure can be assigned to an internal composite source. 
 
Considering black holes as external sources, i.e. not resolved in a physical Hilbert space,
small--scale structure of their interior is a void concept.
Scattering experiments allow to 
extract observables localized outside of the black hole. In particular,
the $1/r$--potential can be recovered for $r>r_{\rm g}$, where $r_{\rm g}\equiv2M_{\mathcal{B}}/M_{\rm P}^2$ 
is the Schwarzschild radius, with $M_{\rm P}$ denoting 
the Planck mass and $M_{\mathcal{B}}$ the black hole mass. Furthermore, resummation 
of tree scattering processes sourced by the external black hole
give rise to geodesic motion in the respective Schwarzschild background.
As explained before this allows for reinterpretation of geometry as being 
emergent from an $S$--matrix defined on flat space--time.

Treating black holes as internal sources anchored in the physical Hilbert space,
their interior quantum structure can be resolved by employing probes of sufficient
virtuality, $-q^2 > r_{\rm g}^{\; -2}$. This can be described in a weakly coupled 
field theory provided $-q^2 < M_{\rm P}^{\; 2}$ holds. 
Notice, that these ideas depart from the semi-classical point of view.
There, the existence of a horizon prohibits an observer outside of the black hole
to get any information about the internal structure of the system.
As explained before, the geometrical concept, is not fundamental 
within our approach.
Rather geometry and thus the existence of a horizon
should be understood as effective phenomena.
On the microscopic level, however, this description should break down
and a resolution of the bound state becomes
possible for an outside observer.

For simplicity, consider an ingoing scalar $\Phi$ outside of the black hole
emitting a graviton with appropriate virtuality, which subsequently gets absorbed 
by another scalar in the black hole's interior. This process is encoded in the linearized 
Einstein--Hilbert action coupled to the energy--momentum tensor of a massless scalar,
$
	\mathcal{T}
	=
	{\rm d}\Phi\otimes {\rm d}\Phi - ({\rm d}\Phi,{\rm d}\Phi) \eta /2 $:

\begin{eqnarray}
S=\int{\rm d}^4x\left[\frac{1}{2} h_{\mu\nu}\epsilon^{\mu\nu}_{\alpha\beta}h^{\alpha\beta}+\frac{1}{M_{\rm{P}}} h_{\mu\nu} \mathcal{T}^{\mu\nu}\right]\; .
\end{eqnarray}

Here $\epsilon^{\mu\nu}_{\alpha\beta}$ is the standard linearized kinetic operator of general relativity
expanded around flat space--time. Note that we can trust this effective action in the kinematic regime discussed
above.

\begin{figure}[h]
\centering
\includegraphics[scale=0.12]{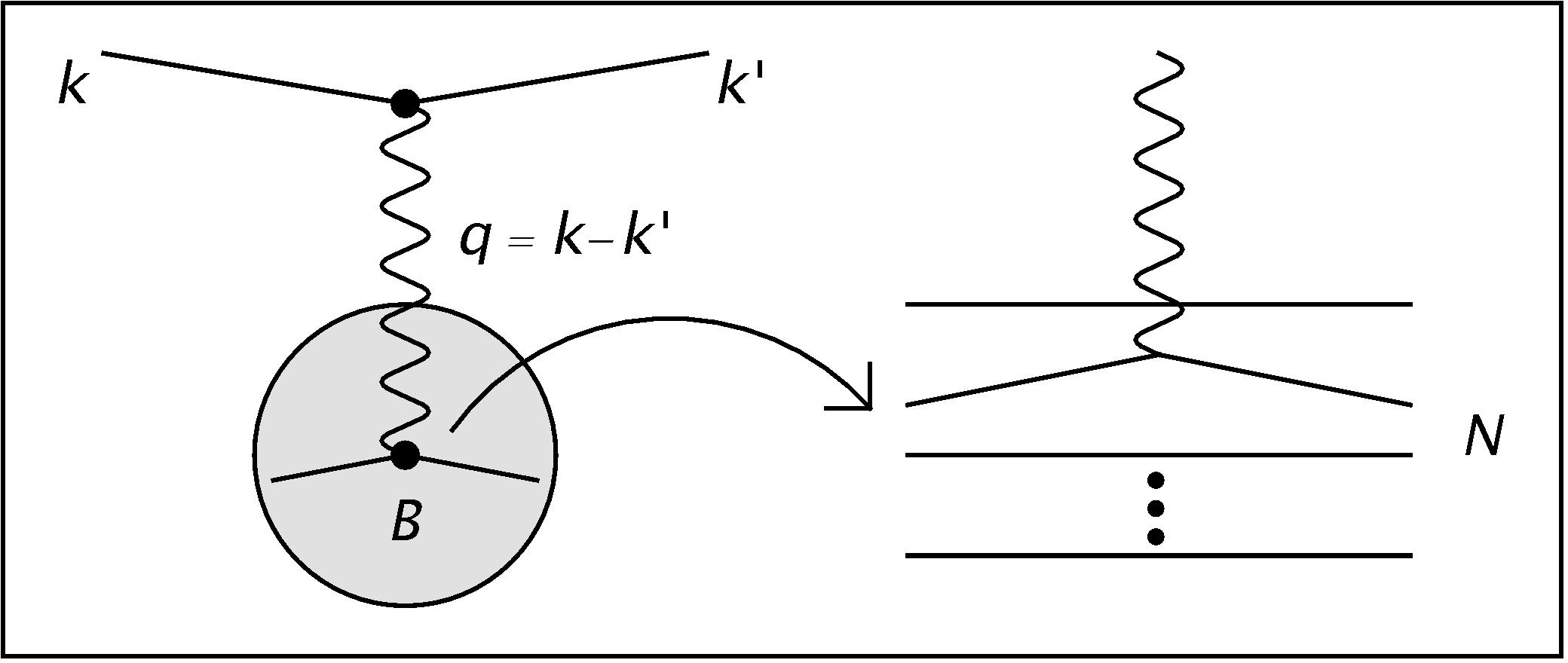}
\caption{Feynman diagram for the scattering of a scalar on a black hole bound state at tree level. 
The wiggly line corresponds to the exchange of a virtual graviton.
On the right hand side, the corresponding absorption is resolved
into the microscopic constituents spectating and participating in the scatter process.}
\label{scatteringendl}
\end{figure}

 Before truncating ingoing and outgoing
emitter legs, the one--graviton exchange amplitude for this process at tree level reads (Figure \ref{scatteringendl})
\begin{eqnarray}
	a^{(2)}(x_1,x_2)
	=
	\tfrac{{\rm i}^2}{M_{\rm P}^{\; 4}} \int {\rm d}^4z_1 {\rm d}^4z_2 \;
	\mathcal{P}^{\mu\nu}(z_1,z_2;x_1,x_2)  \mathcal{N}_{\mu\nu}(z_2),
	\nonumber
\end{eqnarray}	
where $\mathcal{P}$ contains all correlations with respect
to the perturbative vacuum state $|0\rangle$, and $\mathcal{N}$ carries local, non-perturbative
information about the black hole quantum state $|\mathcal{B}\rangle$:
\begin{eqnarray}
	\mathcal{P}^{\mu\nu}
	&=&
	\langle 0|{\rm T} \Phi(x_2) \mathcal{T}_{\alpha\beta}(z_1)\Phi(x_1)|0\rangle \; 
	\Delta^{\alpha\beta\mu\nu}(z_1,z_2) \; ,
	\nonumber \\
	\mathcal{N}_{\mu\nu}
	&=&
	\langle \mathcal{B}^\prime| 
	:\hspace{-0.1cm}\mathcal{T}_{\mu\nu}\hspace{-0.1cm}:(z_2) 
	|\mathcal{B}\rangle
	\; .
\end{eqnarray}	
Here $\Delta$ denotes the free graviton propagator, and $|\mathcal{B}^\prime\rangle$  is 
the black hole quantum state after absorbing the graviton. 
Basically, $\mathcal{P}$ describes space--time events that originate
outside the bound state, while $\mathcal{N}$ 
is localised in its interior. 

Using the auxiliary current description, and provided that the bound state wave function
$B(\mathcal{L})$ has a sufficiently compact support in $\mathcal{L}$--space,
the graviton absorption event can be translated to the origin:
\begin{eqnarray}
	\mathcal{N}(z_2)
	\approx
	{\rm e}^{-{\rm i}(P^\prime - P)\cdot z_2} 
	\langle \mathcal{B}^\prime| :\hspace{-0.1cm}\mathcal{T}\hspace{-0.1cm}:(0)|\mathcal{B}\rangle
	\; ,
\end{eqnarray}	
with $P^\prime$ and $P$ denoting the black hole momentum after and before the graviton absorption,
respectively, around which the corresponding wave function is peaked. 
The evaluation of $\mathcal{P}$ is straightforward. Truncating the ingoing and outgoing 
emitter legs, the one--graviton exchange amplitude becomes  
\begin{eqnarray}
	\langle \mathcal{B}^\prime \Phi^\prime|\mathcal{B}\Phi\rangle^{(2)}
	&&=
	-{\rm i} (2\pi)^4 \delta(k^\prime+P^\prime-k-P)\; 
	\alpha_{\rm g}^{\; 2}
	\nonumber \\
	&&\times \langle \mathcal{B}^\prime| 
	:\hspace{-0.1cm}\mathcal{T}_{\mu\nu}\hspace{-0.1cm}:(0)
	|\mathcal{B}\rangle \;
	\Delta^{\mu\nu\alpha\beta}(k^\prime-k) \;
	G_{\alpha\beta\rho\sigma} \; \tfrac{k^{\prime \rho}k^\sigma}{k^{\prime 0}k^0}
	\; ,
\end{eqnarray}	
where the coupling $\alpha_{\rm g}\equiv 1/(4\pi M_{\rm P}^{\; 2})$ has been introduced,
and \textbf{$G_{\alpha\beta\rho\sigma}=\eta_{\beta(\alpha}\eta_{\rho)\sigma}-\eta_{\beta\sigma}\eta_{\alpha\rho}$} is the Wheeler--DeWitt metric. 

The total cross section $\sigma(\mathcal{B}^\prime\Phi^\prime\leftarrow \mathcal{B}\Phi)$
involves the absolute square of this amplitude and an integration over all intermediate
bound states in the spectrum of the theory. Therefore, the differential cross section 
can be written as
\begin{eqnarray}
	k^{\prime 0} \frac{{\rm d}\sigma}{{\rm d}^{^3}k^\prime} =
	\frac{2}{\mathcal{F}(\Phi)} \;
	\left|\alpha_{\rm g}\Delta(k^\prime-k)\right|^2 \;
	\mathcal{E}^{\alpha\beta\mu\nu}(k,k^\prime) \mathcal{A}_{\alpha\beta\mu\nu}(\mathcal{B};k,k^\prime) \; .
\end{eqnarray}
Here, $\mathcal{F}$ denotes the ingoing flux factor and $\Delta$ the scalar part of
the graviton propagator. The emission tensor $\mathcal{E}$ captures the virtual 
graviton emission outside of the black hole, and the 
absorption tensor $\mathcal{A}$ its subsequent absorption by a black hole constituent.
The emission tensor $\mathcal{E}\equiv Q\otimes Q$ is build from 
\begin{eqnarray}
	Q^{\mu\nu} =
	4\pi^2 \; \Pi^{\mu\nu\alpha\beta}(k^\prime-k)\;  G_{\alpha\beta\rho\sigma} \;
	\tfrac{k^{\prime \rho}k^{\sigma}}{k^{\prime 0}k^{0}}
	\; ,
\end{eqnarray}	
with the graviton polarisation tensor $\Pi_{\mu\nu\alpha\beta}(q)\equiv
\pi_{\mu(\alpha}\pi_{\beta)\nu}-\pi_{\mu\nu}\pi_{\alpha\beta}$, 
where $\pi_{\mu\nu}\equiv\eta_{\mu\nu}-\frac{q_\mu q_\nu}{q^2}$, and $k,k^\prime$ the on--shell
momenta of the ingoing and outgoing scalar emitter, respectively. 
Graviton absorption is described as the energy momentum 
correlation of black hole constituents:
\begin{eqnarray}
	\label{A}
	\mathcal{A}
	=
	\tfrac{1}{2\pi}\int {\rm d}^4x \; {\rm e}^{-{\rm i}(k^\prime-k)\cdot x} 
	\langle\mathcal{B}|\mathcal{T}(x)\otimes\mathcal{T}(0)|\mathcal{B}\rangle
	\; .
\end{eqnarray}	
Clearly, $\mathcal{A}$ contains information about the black hole interior, which 
is not yet resolved in terms of chronologically ordered
subprocesses. For practical calculations, $\mathcal{A}$ 
will be related to the corresponding time ordered amplitude in the next section.

\section{Chronological Ordering}\label{CO}
Given that the graviton absorption tensor is not directly subject to time ordering,
the question arises whether it can be deconstructed into causal correlations.
The method to achieve this is very well-known in the context of scattering
processes on bound states in QCD and will be adapted to the problem at hand 
in the following discussion.

As a first step, let us relate $\mathcal{A}$ to a tensor built from 
$\mathcal{T}(x)\wedge\mathcal{T}(0)$. 
Inserting a complete set of physical states 
in between the energy--momentum tensors
at $x$ and $0$ in (\ref{A}), and making good use of space--time translations, 
we arrive at
\begin{eqnarray}
	\mathcal{A}
	=
	\tfrac{1}{2\pi}\int\hspace{-0.46cm}\sum_\mathcal{B^\prime}
	(2\pi)^4 \delta(q+P-P^\prime)
	\langle B|\mathcal{T}(0)|\mathcal{B}^\prime\rangle \hspace{-0.08cm}
	\langle\mathcal{B}^\prime|\mathcal{T}(0)|\mathcal{B}\rangle
	\; , \nonumber
\end{eqnarray}	
with $q\equiv k-k^\prime$, $P$ and $P^\prime$ denoting the central momenta
of wave--packets corresponding to ingoing and outgoing black hole quantum states, respectively. 
Standard kinematical arguments allow to replace (\ref{A}) with 
\begin{eqnarray}
	\label{A2}
	\mathcal{A}
	=
	\tfrac{1}{2\pi}\int {\rm d}^4x \; {\rm e}^{{\rm i}q\cdot x} 
	\langle\mathcal{B}|[\mathcal{T}(x),\mathcal{T}(0)]|\mathcal{B}\rangle
	\; .
\end{eqnarray}	

The absorption tensor (\ref{A2}) is given by
the absorptive part of the Compton--like amplitude $\mathcal{C}$
for the forward scattering of a virtual graviton off a black hole,
\begin{eqnarray}
	\mathcal{C}
	=
	{\rm i} \int{\rm d}^4 x {\rm e}^{{\rm i}q\cdot x}
	\langle\mathcal{B}|{\rm T} \; \mathcal{T}(x)\otimes\mathcal{T}(0)|\mathcal{B}\rangle
	\; .
\end{eqnarray}	
In order to see this, let us make the discontinuity of $\mathcal{C}$ manifest repeating 
the steps that allowed to extract the kinematical support of $\mathcal{A}$, leading to
\begin{eqnarray}
	\mathcal{C}
	=
	\int\hspace{-0.46cm}\sum_\mathcal{B^\prime}
	\tfrac{(2\pi)^3\delta({\bf P^\prime}-{\bf P}-{\bf q})}{P^{\prime 0}-P^0-q^0-{\rm i}\varepsilon}
	\langle\mathcal{B}|\mathcal{T}(0)|\mathcal{B}^\prime\rangle \hspace{-0.08cm}
	\langle\mathcal{B}^\prime|\mathcal{T}(0)|\mathcal{B}\rangle
	\; .
\end{eqnarray}	
Defining 
Abs $\omega^{-1}\equiv [(\omega-{\rm i\varepsilon})^{-1}-
(\omega+{\rm i\varepsilon})^{-1}]/(2{\rm i})$, it follows that
Abs$(P^{\prime 0}-P^0-q^0-{\rm i}\varepsilon)^{-1}=\pi
\delta(P^{\prime 0}-P^0-q^0)$ and hence, 
\begin{eqnarray}
	\pi\mathcal{A}(\mathcal{B};q)
	=
	{\rm Abs}\; \mathcal{C}(\mathcal{B};q)
	\; ,
\end{eqnarray}
which allows to deconstruct $\mathcal{A}$ in terms of chronologically ordered
correlations.

\section{Constituent representation of $\mathcal{A}$}\label{CR}
In this section we give a physical interpretation of the absorption tensor 
in terms of constituent observables.

The time--ordered product of energy--momentum tensors in $\mathcal{C}$
gives rise to three contributions: 
The first corresponds to maximal connectivity between the tensors,
resulting in a purely perturbative contribution void of any structural
information.
The second represents a disconnected contribution.
Finally, the third contribution allows for perturbative correlations 
between the energy--momentum tensors and, in addition, carries 
structural information. Dropping the contributions void of structural information,
\begin{eqnarray}
	{\rm T} \;\mathcal{T}_{\alpha\beta}(x)\mathcal{T}_{\mu\nu}(0)
	=
	\tfrac{1}{4}
	G_{\alpha\beta}^{ab}G_{\mu\nu}^{mn} \; C_{bm}(x) \; \mathcal{O}_{an}(x,0) \; ,\nonumber
\end{eqnarray}	
where $C(x)\equiv 4 \langle 0|{\rm T} \;{\rm d}\Phi(x)\otimes {\rm d}\Phi(0)|0\rangle$
denotes the correlation with respect to the perturbative vacuum, 
\begin{eqnarray}
	C(x)
	=
	- \frac{2}{\pi^2}
	\frac{x^2\eta - 4 x\otimes x}{(x^2)^3}
\end{eqnarray}	
in free field theory,
and 
$\mathcal{O}(x,0)\equiv \; :\hspace{-0.1cm}{\rm d}\Phi(x)\otimes{\rm d}\Phi(0)\hspace{-0.1cm}:$
is the bi--local operator allowing to extract certain structural information 
when anchored in a quantum bound state.

In order to extract local observables, $\mathcal{O}(x,0)$ has to be expanded in a series
of local operators. In principal this amounts to a Laurent--series expansion of the corresponding
Green function. Let us first focus on its Taylor part:
\begin{eqnarray}
	\Phi(x)
	=
	{\rm exp}\left(x\cdot\partial_z\right) \Phi(z)|_{z=0} \; .
\end{eqnarray}	
The ordinary partial derivative
is appropriate in the free field theory context, otherwise $\mathcal{O}(x,0)$ requires
a gauge invariant completion. 
Then,
\begin{eqnarray}
	\mathcal{O}(x,0)
	=
	\sum_{j=0}^\infty \tfrac{1}{j!} \;  \mathcal{O}^{[j]} (0) \; ,
\end{eqnarray}	
with $\mathcal{O}^{[j]} (0)\equiv \;
:\hspace{-0.1cm}(x\cdot\partial_z)^j{\rm d}\Phi(0)\otimes{\rm d}\Phi(0)\hspace{-0.1cm}:$
and $[j]\equiv j+4$ denotes the mass dimension of the local operator. Note that we suppressed the space--time point $x$
appearing in the directional derivative in order to stress the local character
of the operator expansion.

The fast track to relate $\mathcal{C}$ to constituent observables is to evaluate 
$\mathcal{O}(x,0)$ in a black hole quantum state using the auxiliary current
description. We find for the local operators 
\begin{eqnarray}
	\label{obs}
	\langle \mathcal{B}|\mathcal{O}^{[j]}(0)|\mathcal{B}\rangle
	=
	\kappa \left(x\cdot P\right)^j 
	\langle \mathcal{B}|\Phi(r)\Phi(0)|\mathcal{B}\rangle 
	P\hspace{-0.1cm}\otimes\hspace{-0.1cm}P\, .
\end{eqnarray}	
Here, $\kappa$ denotes a combinatoric factor.
Note that a simple point--split regularisation has been employed 
($r^2\rightarrow 0$). 
The operator appearing on the right hand side
of (\ref{obs}) measures the constituent number density. 

Hence, the 
absorptive part of the forward virtual graviton scattering amplitude $\mathcal{C}$
or, equivalently, the graviton absorption tensor $\mathcal{A}$ can be directly
interpreted in terms of the black hole constituent distribution.

\section{Analytic properties of $\mathcal{C}$}\label{AP}

The Ward--Takahashi identity associated with the underlying gauge symmetry
fixes the tensorial structure $\theta_{\alpha\beta\mu\nu}(q,P)\equiv \Pi_{\alpha\beta}^{ab}\Pi_{\mu\nu}^{mn}\eta_{bm}P_aP_n$ of the amplitude $\mathcal{C}(q,P)$
in accordance with source conservation.
The Laurent--series expansion of $\mathcal{O}(x,0)$ in local operators gives
to leading order up to $\mathcal{O}(q\cdot P/P^2)$
\begin{eqnarray}
	\mathcal{C}(q,P)
	&=&
	\langle \mathcal{B}|\Phi(r)\Phi(0)|\mathcal{B}\rangle  \;
	\theta(q,P)\; \tfrac{\rm -i}{2\pi^2} \sum_{j=-\infty}^\infty C_j(q) u^j
	\; .
\end{eqnarray}
Here, the coefficients $C_j$ are calculable and turn out to be momentum independent,
and the expansion parameter $u\equiv-P^2/q^2\gg 1$. 
Note that this parameter is the analogue of the inverse Bjorken
scaling variable known from deep inelastic scattering.
There is a profound difference between these two parameters, however.
While in standard discussions of deep inelastic scattering in the infinite
momentum frame one makes use of asymptotic freedom, this
is not possible in gravity.
For the problem at hand, however, there is a natural limit
and correspondingly an appropriate expansion parameter.
Namely, considering black holes of large mass
and momentum transfers smaller than $M_{\rm P}$ (which is needed 
in order to trust the perturbative expansion) we are naturally lead 
to the expansion parameter $u$.

The discontinuity of $\mathcal{C}$ for fixed $q^2=-Q^2$
is at
\begin{eqnarray}
	u_*
	=
	\tfrac{M_{\mathcal{B}}}{2(M_{\mathcal{B}}^\prime-M_{\mathcal{B}})} \left(1-\tfrac{M_{\mathcal{B}}^{\prime 2}-M_{\mathcal{B}}^2}{Q^2}\right)
	\gg 1 \; .
\end{eqnarray}	
So $\mathcal{C}$ has an isolated pole at $u_*\gg 1$ and, in particular,
no branch cut in leading order, corresponding to the statement that
$M_{\mathcal{B}}^\prime/M_{\mathcal{B}}-1\approx 0$. Of course, the presence of a branch cut
beyond leading order poses no obstacle. 
On the contrary, it has an evident interpretation in terms 
of intermediate black hole excitations.  

In order to project onto the 
Laurent--coefficients, a path enclosing $[-u_*,u_*]\subset\mathbb{R}$ in the complex $u$ plane 
has to be chosen. 
\begin{figure}[h]
\centering
\includegraphics[scale=0.12]{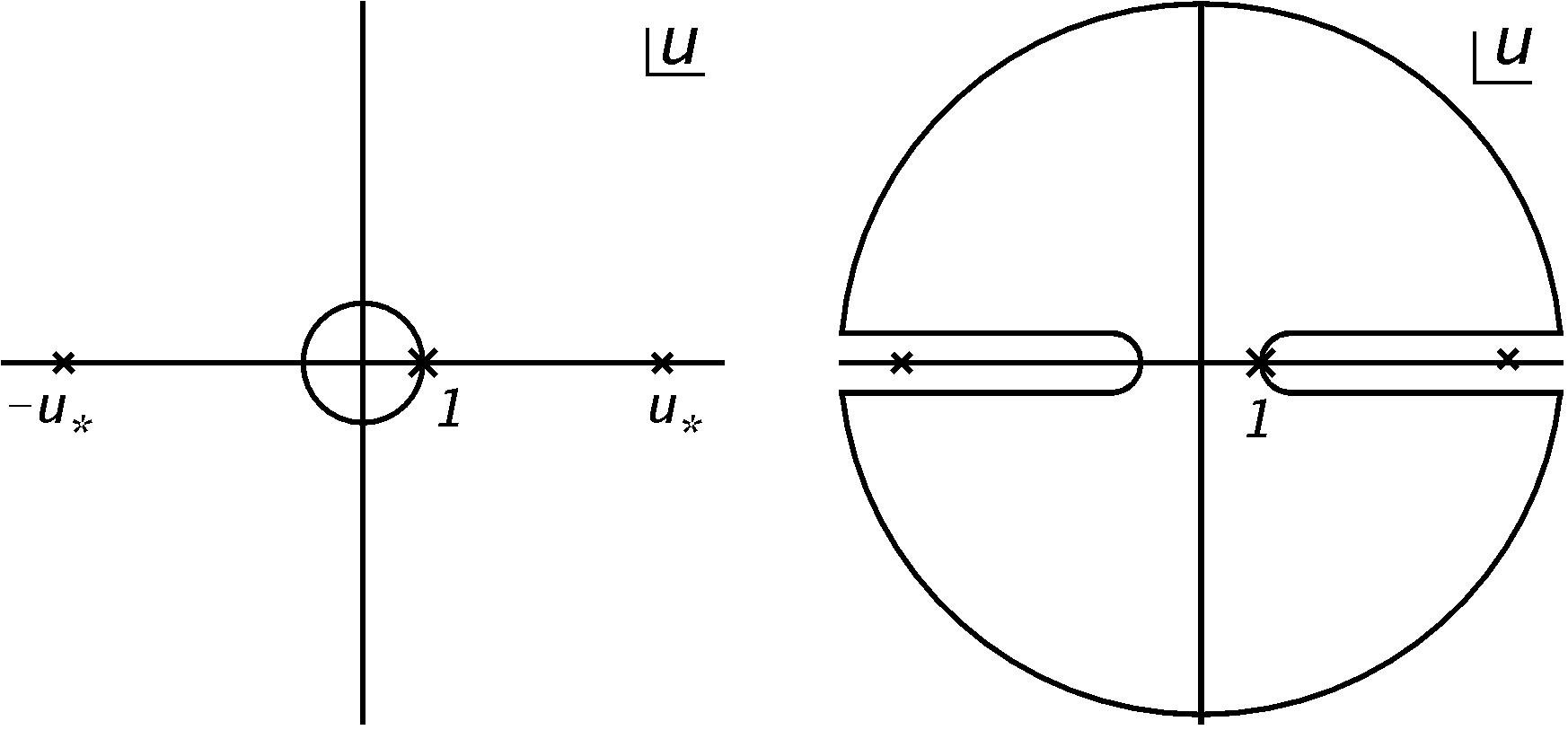}
\caption{Integration contour in the complex $u$-plane. The left figure displays an integration 
contour corresponding to the radius of convergence. In order to relate this to the physical 
$u$-region ($P^2>Q^2$) we perform a contour deformation (right figure). The radius of the circle
is sent to infinity.}
\label{integrationcontourendl}
\end{figure}
This covers the physical $u$ region, while the radius of convergence
of the corresponding Taylor series would only allow for unphysical $u\in[-1,1]$ (see Figure \ref{integrationcontourendl}). 
We find
\begin{eqnarray}
	\int_0^1 {\rm d}\zeta \; \zeta^{k-2} \mathcal{A}(q,P,\zeta) 
	= \tfrac{C_{k-1}}{4\pi^2} \;
	\langle \mathcal{B}|\Phi(r)\Phi(0)|\mathcal{B}\rangle \; \theta(q,P)
	\; ,
\end{eqnarray}	
with $\zeta\equiv 1/u$ denoting the graviton virtuality relative
to the black hole target mass. Hence, all moments of the absorption tensor
with respect to $\zeta$
are directly proportional to the constituent distribution inside the black hole. 
This implies that
${\rm d}\sigma/{\rm d}^3 k^\prime \propto 
\langle \mathcal{B}|\Phi(r)\Phi(0)|\mathcal{B}\rangle$. Thus,
black hole constituent distributions are observables that can be
extracted from scattering experiments.

Although $\langle \mathcal{B}|\Phi(r)\Phi(0)|\mathcal{B}\rangle$ can in practice
not be determined from first principles, we will give a simple toy model
for the wave function in the next section and compute $\mathcal{D}(|{\bf r}|)$.
Requiring that the wave function is localized within the Schwarzschild radius (which
seems to be a sensible assumption), will lead to a qualitative understanding
of the distribution of quanta inside $|\mathcal{B}\rangle$.

In order to allow quantitative statements this means that the distribution should be measured at
some scale $\Lambda \ll M_{\rm{p}}$ where the effective field theory
description is valid. Predicting the cross section at a different scale can then be achieved
by means of renormalization group techniques.
Notice that this procedure is similar to the DGLAP evolution of quark
and gluon distributions within the framework of perturbative QCD.
Renormalization group evolution, however, will be studied in future work.

\section{Constituent distribution at parton level} \label{CD}
In Sections \ref{CR} \& \ref{AP} we presented a constituent interpretation 
for virtual graviton absorption by a black hole quantum bound state.
Central for this interpretation was the constituent 
distribution $\mathcal{D}(r)\equiv \langle \mathcal{B}|\Phi(r)\Phi(0)|\mathcal{B}\rangle$.
As discussed in Section \ref{ACD}, $\mathcal{D}(r)$ can only depend on the spatial
distance $|{\bf r}|$, but not on time.

\begin{figure}[h]
\centering
\includegraphics[scale=0.13]{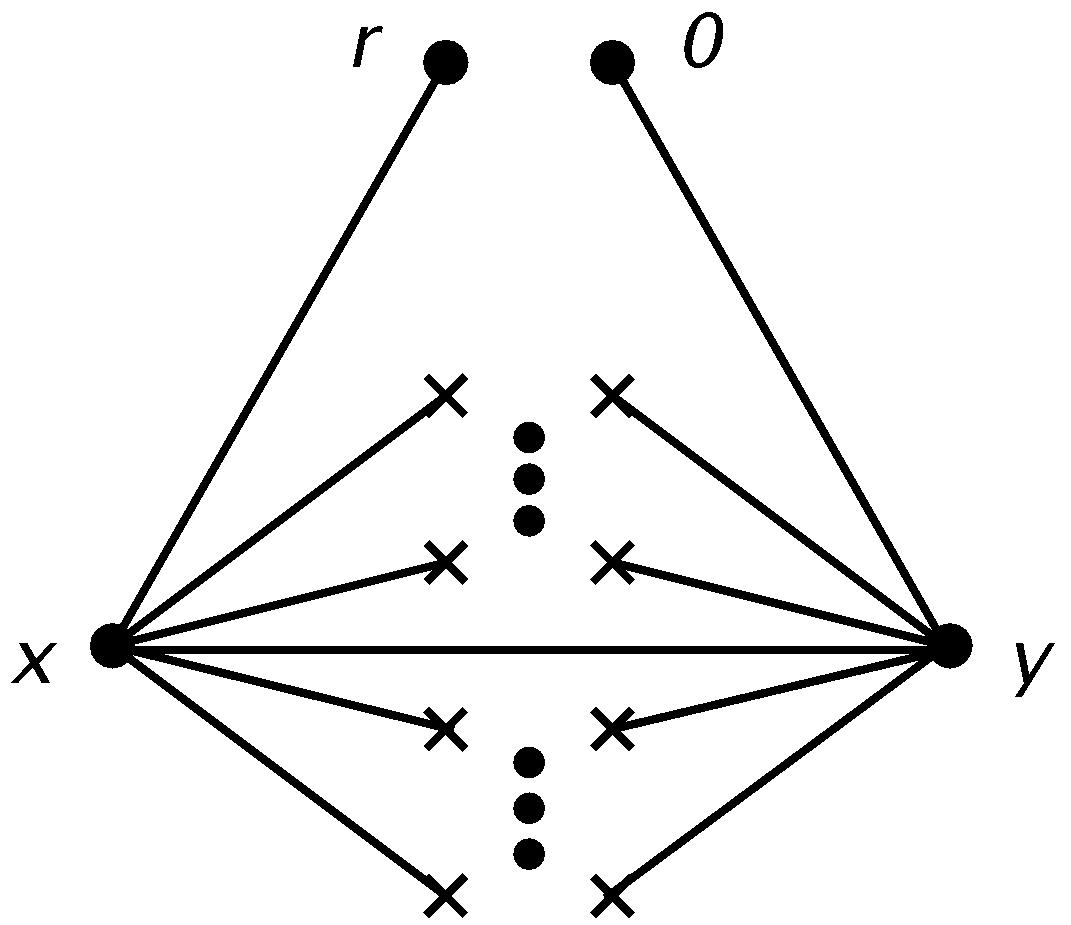}
\caption{This is the only diagram contributing to $ \mathcal{D}(|{\bf r}|)$. Straight lines
represent free propagators, while lines ending in crosses correspond to 
non-perturbative condensation processes.}
\label{numberdensityendl}
\end{figure}

The spatial length scale $|{\bf r}|$ is at the observers disposal.
It can be interpreted as the necessarily finite spatial extent of an apparatus 
that emits a $\Phi$ quantum at one end and subsequently absorbs it at the other end. 
In between emission and absorption the quantum probes the medium in which
the apparatus has been submerged, in our case the black hole interior.
At the partonic level, no individual interactions between the probe and black hole
constituents take place, therefore the only relevant 
scale remains $|{\bf r}|$. Effectively, then, there is only the correlation
across the apparatus between emission and absorption events, which scales as $|{\bf r}|^{-2}$.
The light--cone distribution of black hole constituents has been
calculated in \cite{Hofmann} at the parton level and to leading order in $1/N$.
As explained in \cite{Hofmann} these $1/N$--corrections already arise at the level of combinatorics
associated to the diagrams that need to be computed. Note that these corrections are in accordance
with the ideas of \cite{Dvali}.
\begin{eqnarray}
\mathcal{D}(|{\bf r}|)=2\Big(\frac{N}{M_{\mathcal{B}}}\Big)^4\Gamma_{\mathcal{B}}^{-2}\langle\Phi^{2(N-2)}\rangle
\frac{1}{4\pi^2 |{\bf r}|^2}\int d^3P \cos({\bf P r}/2)|\mathcal{B}({\bf P},\mathcal{Q})|^2,
\label{distribution}
\end{eqnarray}
where $\langle\Phi^{2(N-2)}\rangle$ is a condensate parametrizing the non--perturbative 
vacuum structure inside a black hole quantum bound state. 
At the diagrammatic level, $\mathcal{D}(|{\bf r}|)$ can be represented by 
Figure \ref{numberdensityendl}. Note that gauge corrections to (\ref{distribution})
can be calculated following \cite{Hofmann}.

Even in the absence of gauge interactions, $N$
carries non--perturbative information via its dependence on $\Gamma_\mathcal{B}$
and, in addition, its dependence on $\Phi$ condensates, see \cite{Hofmann}. 
The latter dependence deserves elaboration. It can be traced back to the fact 
that $N\gg 1$ for black holes, implying minimal
connectivity between the space--time events at which the auxiliary currents 
are operative. 
At the level of $\mathcal{D}(|{\bf r}|)$, this can be
seen as follows. The constituent distribution is generated by a four--point
correlator, where two space--time points are associated with the read--in
events (auxiliary current locations) and one point--split for localising
an apparatus of finite extent, consisting of an emitter and an absorber. 
The measurement process requires altogether six $\Phi$ fields at
four space--time locations. The vast majority of fields composing the 
auxiliary currents has two options. Either they enhance the connectivity
between the currents locations, or they condense. Condensation of $\Phi$ quanta
turns out to be the favoured option in the
so--called double scaling limit, 
$N, M_{\mathcal{B}}\rightarrow \infty$ and 
$N/M_{\mathcal{B}}=$const., where $M_{\mathcal{B}}$
denotes the bound state's mass.

Violations of this limit are not exponentially suppressed, but of order 
$1/N\ll 1$, indicating the essential quantum character 
of black holes. Since the cross section is given in terms of the number density, 
these corrections are in principle measurable in scattering experiments.

Black holes naturally are large--$N$ quantum bound states. 
If operative as sources, they trivialize (planar dominance \cite{tHooft,Witten}) the underlying quantum theory
of constituent fields and represent an unrivalled realisation of the 
large--$N$ limit in nature.

\begin{figure}[!h]
     \centering
     \includegraphics[scale=0.45]{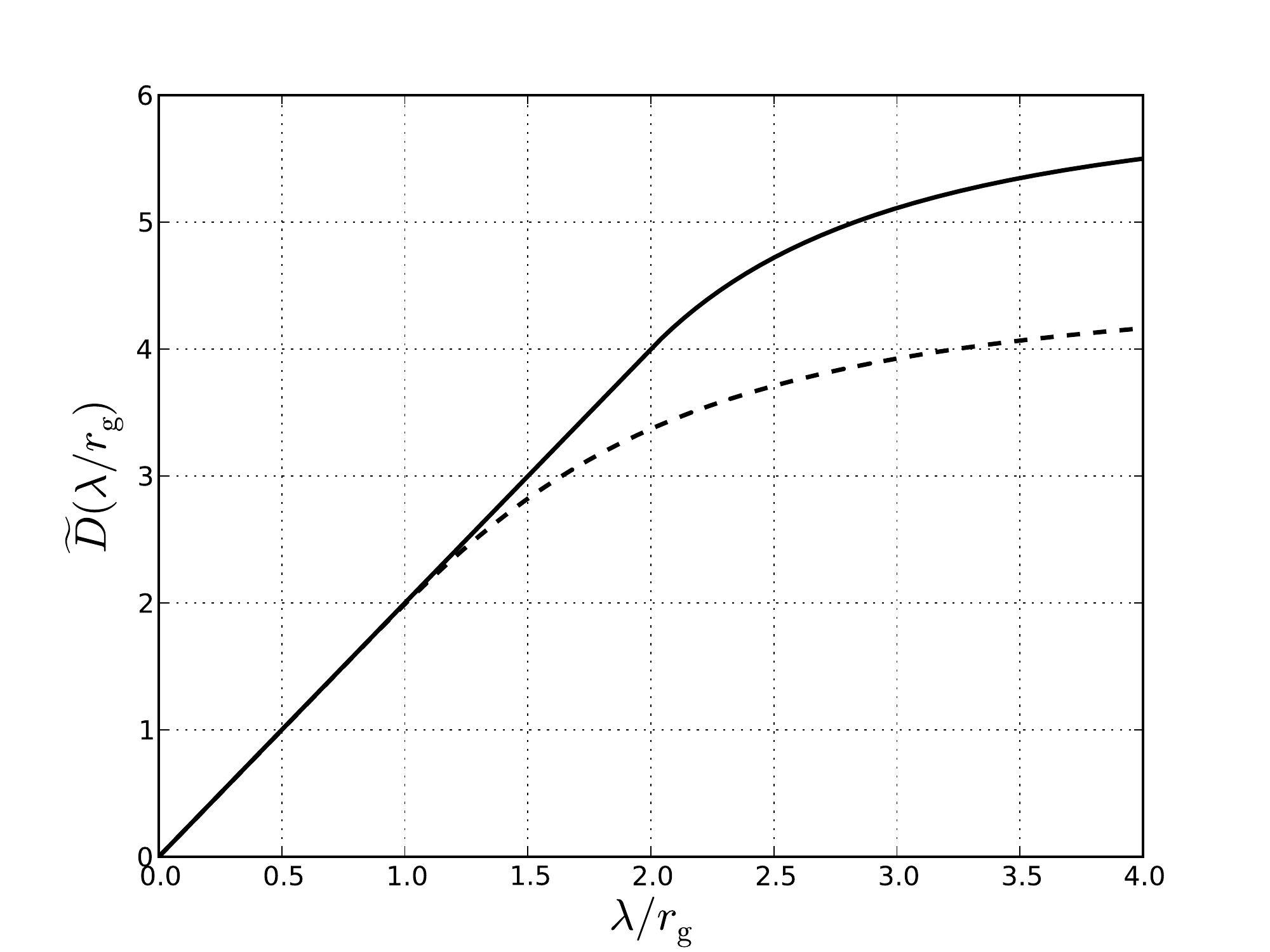}
      \caption{Parton distribution as a function of the wavelength 
		for a gaussian wave function of variance $r_{\rm g}$ (dashed line) and a Heaviside
      profile (solid line).
      Here $\tilde{\mathcal{D}}(\lambda/r_{\rm g})$ is normalised to 
	  $\frac{r_{\rm g}}{2}(N/M)^4\Gamma_{\mathcal{B}}^{-2}\langle\Phi^{2(N-2)}\rangle$. 
	  The distribution is plotted only for wavelengths $\lambda<4r_{\rm g}$ since 
      the condensates are not supported outside the black hole. 
	  By construction our analysis is valid only 
      up to $\lambda=2r_{\rm g}$. }
          \label{partdistbhl}
    \end{figure}

Before concluding this section, let us calculate the constituent 
distribution $\mathcal{D}(|{\bf r}|)$ assuming
a Gaussian wave function $\mathcal{B}({\bf P},\mathcal{Q})$ for the black hole
peaked around $M_{\mathcal{B}}$ with a standard deviation given by $1/r_{\rm g}$.
This choice reflects those features of the a priori unknown black hole state 
that are relevant for the qualitative behaviour of the constituent distribution. 
For instance, the non--perturbative ground state 
has compact support characterised by the size of the bound state itself.
In Figure \ref{partdistbhl} we show the constituent distribution in wavelength space,
$\widetilde{D}(\lambda)\propto \lambda \erf(2r_{\rm g}/\lambda)$, where
$\erf$ denotes the error function and $\lambda=1/|{\bf k}|$. Here $|{\bf k}|$ is the absolute value of the constituent three--momentum.
As can be seen, black hole constituents favour to
occupy long wavelength modes. In other words, black hole interiors are dominated
by soft physics in accordance with the postulates of reference \cite{Dvali}.

\section{Discussion \& Summary}\label{Discussion}
Black holes are perhaps the most celebrated solutions of general relativity.
Within our framework they are considered as bound states
of quantum constituents on flat space--time with physical radius $r_{\rm g}$. 

We discussed the representation of bound states in terms of currents in detail.
Subsequently we specialised to spherically symmetric gravitating sources
including black hole quantum bound states.

A quantum theory of black hole constituents allows to extract structural information
from the associated quantum bound state. We showed that the process of virtual graviton
absorption by a black hole is directly related to the constituent distribution 
inside the black hole. In particular, we gave a precise prediction for the
differential scattering cross section of massless scalars on a black hole
in terms of microscopic degrees of freedom constituting the bound state. 
Hence the constituent distribution is a faithful observable ---
it can be defined using a gauge invariant operator and, in addition, a scattering process
can be specified allowing its measurement. Thus, in contrast to the standard
lore, within our framework an outside observer in principle has access to the 
internal structure of a black hole.

We discussed the physics underlying graviton absorption by black holes. The quantum
bound state proposal employed here is based on individually weakly coupled constituents 
immersed in a non--trivial medium. Constituent condensation is supported by 
a non--perturbative ground state $|\Omega\rangle$ and 
by the large--$N$ character of black holes. The quantum bound states
associated with black holes can be generated by operating with an appropriate 
current $\mathcal{J}$ on $|\Omega\rangle$. These sources 
trivialise the underlying field theory and allow to consider black holes
as the simplest realisations of large--$N$ bound states in nature.
Finite $N$ effects can (and should) be studied, since they are not
exponentially suppressed, proving that the bound state construction is truly quantum,
and, consequently, that black holes are essentially beyond a semi--classical description. 
Furthermore, higher order radiative corrections to the scattering process
leading to evolution equations for the distribution function should be considered
in the future.

\acknowledgments
It is a great pleasure to thank 
Dennis Dietrich, Gia Dvali, Cesar Gomez, Claudius Krause, Florian Niedermann and Andreas Sch\"afer for rocking discussions.
The work of LG, SM and TR was supported by the 
International Max Planck Research School on Elementary Particle Physics.
The work of SH was supported by
the DFG cluster of excellence `Origin and Structure of
the Universe' and by TRR 33 `The Dark Universe'.


\end{document}